\documentclass[submission, Phys]{SciPost}
\usepackage[english]{babel}
\usepackage{amsthm}
\usepackage{amsmath}
\usepackage{amssymb}
\usepackage{yhmath}
\titlespacing*{\section}{0pt}{0.6\baselineskip}{0.6\baselineskip}
\usepackage[table]{xcolor}
\usepackage[mathscr]{euscript}
\usepackage{titlesec}
\theoremstyle{definition}

\theoremstyle{remark}

\usepackage{comment}

\begin{document}

\begin{center}{\Large \textbf{
Thermodynamic Remnants in Black-hole Evaporation
}}\end{center}

\begin{center}
I. Arraut\textsuperscript{1$\dagger$}, A. K. Mehta\textsuperscript{2}
\end{center}

\begin{center}
{\bf 1} Institute of Data Engineering and Sciences (IDEAS)\\
and Institute of Science and Environment (ISE)\\
University of Saint Joseph,\\
Estrada Marginal da Ilha Verde, 14-17, Macao, China
\\
ivan.arraut@usj.edu.mo\\
{\bf 2}Department of Physics, Kyunghee University, Seoul, Republic of Korea
\\
abhishek.mehta@khu.ac.kr
\end{center}
\begin{center}
\today
\end{center}

\section*{Abstract}
{\bf  We show that black-hole remnant scenario naturally arises in the original computations of Hawking without extra assumptions.

\let\thefootnote\relax\footnotetext{$\dagger$ Corresponding Author}
}

\vspace{10pt}
\noindent\rule{\textwidth}{1pt}
\tableofcontents\thispagestyle{fancy}
\noindent\rule{\textwidth}{1pt}
\vspace{10pt}

\section{Introduction}

Classical black-holes are dense, compact objects with an event horizon that causally separates the interior of the black-hole from its exterior which implies any particle that goes beyond the event horizon is irretrievably lost to any observer outside the horizon \cite{1}. Therefore, classically, black-holes cannot radiate. However, Hawking \cite{Hawking:1975vcx} in his seminal paper demonstrated that due to quantum mechanical effects black-holes do indeed radiate and, as a process, evaporate in time which is called Hawking radiation. Hawking further demonstrated that the spectrum of black-hole radiation is thermal and has a characteristic temperature which is inversely proportional to the black-hole mass $T_H = \frac{1}{8\pi M}$ \cite{Hawking:1975vcx}. Notice that the Hawking temperature has a divergence at $M\to 0$ which can be naturally remedied by proposing remnant scenarios as black-hole endstages \cite{Ong:2015dja, Ong:2024dnr}. According to the proposal, when a black-hole shrinks, at some scale quantum gravity effects kick in \cite{Bonano, Arraut:2008hc, Nowakowski:2009ha, Arraut:2009an,  Bianchi:2018mml} and prevent further evaporation of the black-hole, then, whatever is left of the black-hole is called the remnant. Remnant scenarios emerge naturally in various Generalized Uncertainty Principles (GUP) \cite{Custodio:2003jp, Nozari:2005ah, Arraut:2008hc, Nowakowski:2009ha, sakalli2025zitterbewegung}, quantum gravity theories \cite{Bonano, Bianchi:2018mml, rovelli2018small} and non-commutative models \cite{Arraut:2009an, Mehdipour_2016}, but the physical existence of these remnants are disputed via information theoretic arguments in the context of black-hole information paradox \cite{Ong:2024dnr}. However, such arguments only suggest that it is not guaranteed that remnants, despite its existence, can solve the black-hole information paradox \cite{Bekenstein:1993dz}. It doesn't directly imply that remnants are unphysical. Therefore, by demonstrating remnants as a part of the original computations of Unruh \cite{Unruh:1974bw}, Hawking \cite{Hawking:1975vcx} and DeWitt \cite{dewitt1975quantum}, we demonstrate remnant endstage  scenarios as a fundamental and physically viable aspect of black-hole evaporation without loss of generality rather than a consequence of specific assumptions about Planck's scale physics irrespective of its utility in resolving the black-hole information paradox. Hence, in this paper, we demonstrate that remnant scenarios naturally emerge within the original computation of Hawking itself without extra assumptions regarding Planck's scale physics.

\section{Remnants in thermodynamic black-hole evaporation}
\label{rem_con}
We start with the final result of the Hawking's computation \cite{Unruh:1974bw, Hawking:1975vcx, dewitt1975quantum}

\begin{align}
\frac{\mathrm{d} M}{\mathrm{~d} t}&=\lim _{r \rightarrow \infty} \int_0^\pi \mathrm{d} \theta \int_0^{2 \pi} \mathrm{~d} \phi\left\langle T_{r t}\right\rangle\notag\\&=-\frac{1}{2 \pi} \sum_{l=0}^{\infty} \int_0^{\infty}(2 l+1)\left|B_l(p)\right|^2 \frac{p}{\mathrm{e}^{8\pi M p }-1} \mathrm{~d} p
\equiv -\frac{1}{2\pi}\int_{0}^{\infty}\frac{pF(p)}{\mathrm{e}^{8\pi M p}-1} \mathrm{~d} p \label{mlr}
\end{align}
where in the last equality we have defined a combined greybody factor
\begin{align}
    F(p) \equiv \sum_{l=0}^{\infty} (2 l+1)\left|B_l(p)\right|^2
\end{align}
However, dealing with an infinite sum and an integral together is computationally inconvenient. Therefore, we will represent the above sum via an integral. This technique is called resummation in the literature. Let $f$ be a Laplace resummation of $F$
\begin{align}
    F(p) = 8\pi\int^{\infty}_0 d\mu ~f(\mu) e^{-8\pi \mu p}\label{lpT}
\end{align}
such that we have
\begin{align}
    \frac{\mathrm{d} M}{\mathrm{~d} t} = -4\int_0^{\infty}\int_0^{\infty}\frac{pf(\mu) e^{-8\pi \mu p}}{\mathrm{e}^{8\pi Mp}-1} \mathrm{~d} p\mathrm{~d} \mu
\end{align}
The integral over $p$ in the RHS can be computed to give
\begin{align}
    \int_0^{\infty}\frac{e^{-8\pi \mu p}}{\mathrm{e}^{8\pi Mp}-1} p \mathrm{~d} p &= \int_0^{\infty}\frac{e^{-8\pi \mu p}}{1-\mathrm{e}^{-8\pi Mp}} \mathrm{e}^{-8\pi Mp}~p\mathrm{~d} p \notag\\&= \frac{1}{64\pi^2 M^2}\int_0^{\infty}\frac{e^{-\left(\mu/M+1\right)t}}{1-e^{-t}} t\mathrm{~d} t = \frac{1}{64\pi M^2}\psi'\left(\frac{\mu}{M}+1\right)
\end{align}
where in the above we made a change of variables $8\pi M \to t$ and where $\psi$ is the digamma function. In the above, we utilized the integral representation of the digamma function by Gauss\cite[\href{https://dlmf.nist.gov/5.9E12}{(5.9.12)}]{NIST:DLMF}
\begin{align}
&\psi(z)=\int_0^{\infty}\left(\frac{e^{-t}}{t}-\frac{e^{-z t}}{1-e^{-t}}\right) d t\\
    &\implies\psi'(z) = \int_0^{\infty}\frac{e^{-z t}}{1-e^{-t}} t ~d t
\end{align}
Therefore, we obtain an alternative rendition of Hawking's \cite{Hawking:1975vcx}, Unruh's \cite{Unruh:1974bw} and DeWitt's \cite{dewitt1975quantum} mass loss rate in Eq. (\ref{mlr})
\begin{align}
\frac{\mathrm{d} M}{\mathrm{~d} t} = -\frac{1}{16\pi^2 M^2} \int_0^{\infty} d\mu f(\mu)\psi'\left(1+\frac{\mu}{M}\right)\label{nontris}
\end{align}
We now look at the large mass limit of the above by using the following
\begin{align}
    &\lim_{M \to \infty}\psi'\left(1+\frac{\mu}{M}\right) = \psi'(1) = \zeta(2) = \frac{\pi^2}{6}\label{lowlim}
\end{align}
which leads to
\begin{align}
    \frac{\mathrm{d} M}{\mathrm{~d} t} \overset{\text{large}~ M}{=}  -\frac{1}{96M^2}\int_0^{\infty} d\mu f(\mu)
\end{align}
In the large mass limit, the black-hole behaves like a perfect black-body as expected. Now, we look at the $M\to 0$ limit of Eq. (\ref{nontris}). In the $M\to 0$ limit, the spacetime metric becomes flat and the massless scalar theory in Minkowski space is a Conformal Field Theory (CFT) and therefore, has scale invariance. Hence, in the $M\to 0$ limit, $f(\mu)$ has the following transformation under scaling\footnote{See Appendix \ref{FLL}}
    \begin{align}
    f(\lambda\mu) = \lambda f(\mu)
\end{align}
due to which Eq. (\ref{nontris}) becomes
\begin{align}
    \frac{\mathrm{d} M}{\mathrm{~d} t} \overset{\text{small}~M}= -\frac{M^2}{16\pi^2 } \int_0^{\infty} d\mu' f(\mu')\psi'\left(1+M\mu'\right) \overset{M\to 0}{=} 0 \quad \mu' = \frac{\mu}{M^2}\label{nontris2}
\end{align}
where in the above we worked with $\lambda = M^{-2}$. Solving the above gives 
\begin{align}
    M(t) \overset{\text{small}~M}{=} M_r 
\end{align}
where $M_r$ is the integration constant that precisely corresponds to the remnant mass. This shows that remnant scenarios seamlessly connect with the original computations of Hawking, Unruh and DeWitt without extra assumptions regarding Planck's scale physics!  In fact, the above shows that remnant scenarios simply arise due to the conformal symmetries that emerge as $M\to 0$ instead of Quantum Gravity effects. In the above, the exchange of limits and integration is justified because $\psi'(1+x) \leq \psi'(1)$ for $x > 0$ which makes the integrands in Eq. (\ref{nontris}) and Eq. (\ref{nontris2}) Lebesgue integrable and satisfy the Lebesgue's dominated convergence theorem \cite{evans2025measure}. Even though classically, $M\to 0$ limit represents the vanishing of the horizon, quantum mechanically there is still a remnant horizon of radius $2M_r$ where $M_r$ is the mass of the remnant. This is not surprising because classically the black-hole mass $M$ is also time-independent as well and yet quantum mechanically, $M$ is time-dependent due to Eq. (1) as shown by Hawking. This is just another one of those aspects where classical notions of a system doesn't translate to its quantum mechanical counterpart. 
\section{Fermionization of Black-hole Event Horizon}
\label{rem_MMo}
\begin{figure}[h!]
    \centering
    \includegraphics[width=1\linewidth]{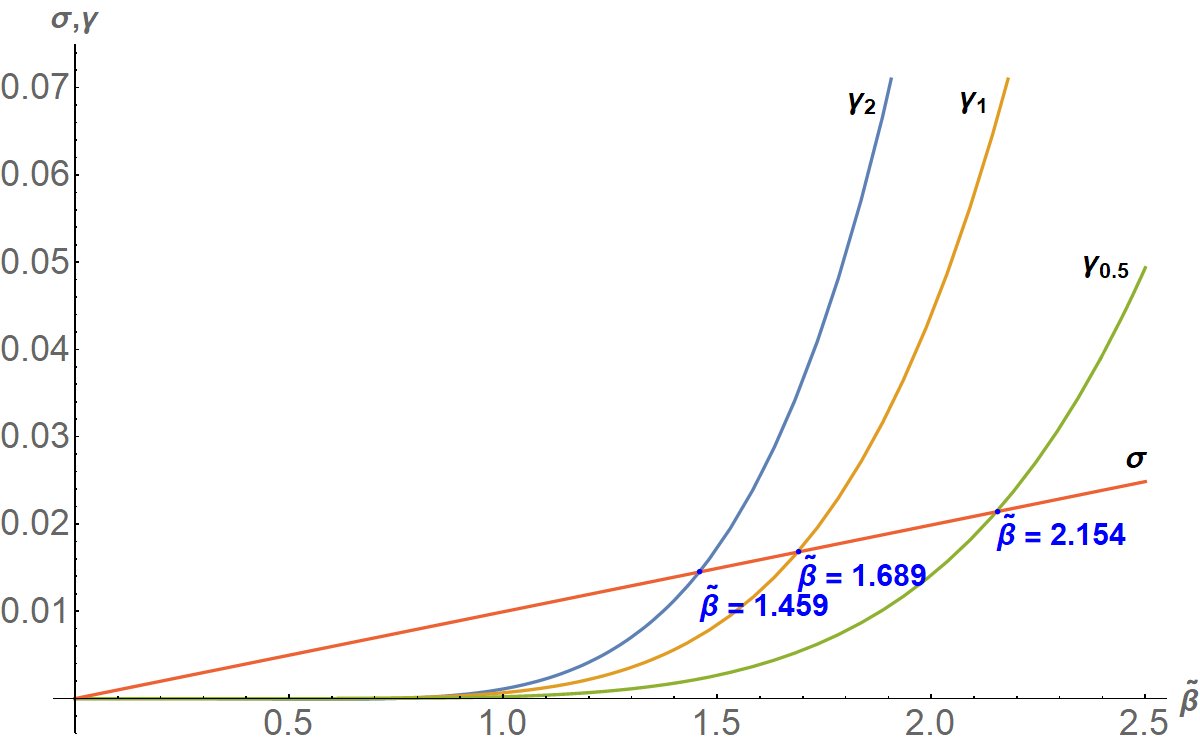}
    \caption{A plot of $\gamma_{\nu} = \frac{\widetilde{\beta}^3}{32\pi^2}\left[\frac{\widetilde{\beta}^2}{2}\ln\widetilde{\beta}~\nu^2-\ln\sqrt{2\pi} +\operatorname{Re}\psi^{(-2)}\left(1+i\nu\widetilde{\beta}\right)\right]$, which represents the pressure $\gamma$ as a function of $\widetilde{\beta}$ for a given Fermi energy $\nu$, and surface tension $\sigma = \frac{\widetilde{\beta}}{32\pi}$. Since, $\widetilde{\beta}$ represents inverse of the black-hole mass, the graph plots the surface tension $\sigma$ and surface pressure $\gamma_\nu$ of the event-horizon for a Fermi level $\nu$ as a function of decreasing black-hole mass. The subscripts to $\gamma$ in each plot represents the value of the Fermi level $\nu = 0.5, 1, 2$. The intersection of $\sigma$ and $\gamma_{\nu}$ computes the $\widetilde{\beta}$ that solves Eq. (\ref{rem_eq}) i.e. the remnant mass $M_r$ for a given $\nu$. Below we show that the stability condition in Eq. (\ref{stabcon}) for each of the remnant mass computed above is indeed satisfied.\\}

\begin{tabular}{||c c c||} 
 \hline
 $\nu$ & $M_r = 1/\widetilde{\beta}$ & $\sigma'(M_r)-\gamma'(M_r)$ \\ [0.5ex] 
 \hline\hline
  0.5 & 0.464 & 0.214  \\ 
 \hline
  1 & 0.592 & 0.135  \\
 \hline
  2 & 0.685 & 0.109  \\ [1ex] 
 \hline
\end{tabular}
    \label{rem_sol}
\end{figure}

In this section, we will describe the full microscopic phenomenology that leads to creation of remnants within the original computation of Hawkings. Consider now a Fourier resummation of $F$ given by
\begin{align}
    F(p) = 8\pi \int^{\infty}_{-\infty} d\mu ~\varphi(\mu) \sin(8\pi \mu p)+8\pi \int^{\infty}_{-\infty} d\mu ~\chi(\mu)\cos(8\pi \mu p)\label{FT}
\end{align}
such that we have
\begin{align}
    \frac{\mathrm{d} M}{\mathrm{~d} t} &= -4\int_{-\infty}^{\infty}\int_0^{\infty}\frac{p\varphi(\mu) \sin(8\pi\mu p)}{\mathrm{e}^{8\pi Mp}-1} \mathrm{~d} p\mathrm{~d} \mu -4\int_{-\infty}^{\infty}\int_0^{\infty}\frac{p\chi(\mu) \cos(8\pi\mu p)}{\mathrm{e}^{8\pi Mp}-1} \mathrm{~d} p\mathrm{~d} \mu\notag\\&= \frac{1}{16\pi^2M^2} \int_{-\infty}^{\infty}\mathrm{~d} \mu~\varphi(\mu)\operatorname{Im}\psi'\left(1+\frac{i\mu}{M}\right)-\frac{1}{16\pi^2M^2}\int_{-\infty}^{\infty}\mathrm{~d} \mu~\chi(\mu)\operatorname{Re}\psi'\left(1+\frac{i\mu}{M}\right)\notag\\
    &=-\frac{1}{16\pi^2M} \int_{-\infty}^{\infty}\mathrm{~d} \mu~\varphi(\mu)\frac{\partial}{\partial\mu} \operatorname{Re}\psi\left(1+\frac{i\mu}{M}\right)-\frac{1}{16\pi^2M} \int_{-\infty}^{\infty}\mathrm{~d} \mu~\chi(\mu)\frac{\partial}{\partial\mu} \operatorname{Im}\psi\left(1+\frac{i\mu}{M}\right) \label{int2}
\end{align}
The above reproduces the same behaviour as the previous section in the various limits which can be ascertained by making use of\footnote{See Appendix \ref{FLL}}
\begin{align}
    &\lim_{M\to \infty}\operatorname{Im}\psi'\left(1+\frac{i\mu}{M}\right) = 0 \quad \lim_{M\to \infty}\operatorname{Re}\psi'\left(1+\frac{i\mu}{M}\right) = \psi'(1)= \zeta(2)\notag\\
    & \varphi(\lambda \mu) =\lambda\varphi(\mu) \quad    \chi(\lambda\mu) =\lambda\chi(\mu) ~~\text{as}~~M\to 0
\end{align}
Notice that in the first integral of the last line of Eq. (\ref{int2}), we have
\begin{align}
    \frac{\partial\rho(\mu)}{\partial\mu} \equiv -\frac{1}{2\pi}\frac{\partial}{\partial\mu} \operatorname{Re}\psi\left(1+\frac{i\mu}{M}\right)
\end{align}
This is precisely the derivative of the density of state $\rho(\mu)$ of $N\times N$ matrix model\cite{gross1990nonperturbative} with a constant Hamiltonian shift. The density of state $\rho(\mu)$ is then given by\cite{gross1990nonperturbative}
\begin{align}
    \rho(\mu) = \frac{1}{2\pi}\left[\ln\widetilde{\beta}-\operatorname{Re}\psi\left(1+i\widetilde{\beta}\mu\right)\right]
\end{align}
such that 
\begin{align}
\widetilde{\beta} = M^{-1} 
\end{align}
where $\widetilde{\beta}$ is a parameter in the matrix models \footnote{For our purposes, we think of $N \times N$ matrix models as a quantum mechanical theory of matrices rather than as limits of certain string theories.}. Using the above identification, we can rewrite Eq. (\ref{int2}) as
\begin{align}
    \frac{\mathrm{d} M}{\mathrm{~d} t} = \frac{1}{8\pi M} \int_{-\infty}^{\infty}\mathrm{~d} \mu~\varphi(\mu)\frac{\partial\rho(\mu)}{\partial\mu}+\cdots\label{int3}
\end{align}
We can further rewrite the above after integration by parts as
\begin{align}
    \frac{\mathrm{d} A}{\mathrm{~d} t} = -4\int_{-\infty}^{\infty}\mathrm{~d} \mu~\varphi'(\mu)\rho(\mu)+\cdots \label{FGRule}
\end{align}
Notice that the RHS looks like the Fermi's Golden rule for transition rates in an $N\times N$ matrix model where $\varphi'(\mu)$ plays the role of transition probability between the quantum states of the $N\times N$ matrix model. To see this, we quote the black-hole evaporation rate computed via the Fermi's golden rule in a generic quantum mechanical model of Hawking radiation which is characterized by some interacting Hamiltonian $\hat{H}_{int}$ \cite{Rakic:2025svg}
\begin{align}
&\left.\frac{d N}{d t}\right|_{\text {quantum-corrected }}=\int d \omega \int d E_f \rho\left(E_f\right) \Gamma_{i \rightarrow f}\notag\\
&\Gamma_{i \rightarrow f}=2 \pi|\langle E_f, \omega| \hat{H}_{\mathrm{int}}| E_i\rangle|^ 2 \delta\left(E_f-E_i-\omega\right)\label{FGRule2}
\end{align}
where the following correspondence between Eq. (\ref{FGRule}) and Eq. (\ref{FGRule2}) holds 
\begin{align}
\mu \leftrightarrow E_f \quad \varphi'(\mu) \leftrightarrow \int d\omega~ \Gamma_{i\to f} 
\end{align}
which also gives a physical interpretation to $\varphi(\mu)$ as encoding the interaction Hamiltonian. Fermi's Golden Rule is routinely utilized in literature to model black-hole horizon dynamics \cite{Bai:2023hpd, Lin:2025wof, Biggs:2025nzs}. The interacting Hamiltonian $\hat{H}_{int}$ characterizes the underlying quantum mechanical model for the dynamics and also defines the black-hole states $|E_i\rangle$.  Since, $N \times N$ matrix models are also fermionic theories \cite{Kazakov:1988ch}, this represents the fermionization of the black-hole event horizon! In other words, the horizon subsystem can be described by a dual theory of $2D$ fermionic gas.  This now allows us to define the grand potential $\Omega$ for the horizon subsystem as follows \cite{Bocquet_2019}
\begin{align}
    \Omega = -\frac{1}{\beta}\int_0^{\infty}\rho(e)\ln[1+\mathrm{e}^{\beta\left(\varepsilon-e\right)}]  \mathrm{d} e \equiv -\frac{1}{\beta}\ln\mathcal{Z} \quad e = \widetilde{\beta}\mu
\end{align}
where $\varepsilon$ is the chemical potential. We define
\begin{align}
&\Phi(e) = \frac{1}{2\pi}\left[\ln\widetilde{\beta} ~e-\operatorname{Im}\ln\Gamma\left(1+ie\right)\right] \quad \rho(e) \equiv \Phi'(e)
\end{align}
From which one can obtain
\begin{align}
    &\frac{\partial\Omega}{\partial \varepsilon}=-N=-\int_0^{\infty} \rho(e) \frac{1}{1+\mathrm{e}^{\beta\left(e-\varepsilon\right)}} \mathrm{d} e\notag\\
    &\implies\frac{N}{\widetilde{\beta}} = \int_0^{\infty} \rho(\mu) \frac{1}{1+\mathrm{e}^{\beta\widetilde{\beta}\left(\mu-\nu\right)}} \mathrm{d} \mu \quad e \equiv \widetilde{\beta}\mu\quad \varepsilon \equiv \widetilde{\beta}\nu\\
    &\Omega= -\int_0^{\infty}\Phi(e)\frac{1}{1+e^{\beta(e-\varepsilon)}}  \mathrm{d} e = -\widetilde{\beta}\int_0^{\infty}\Phi(\mu)\frac{1}{1+\mathrm{e}^{\beta\widetilde{\beta}\left(\mu-\nu\right)}}  \mathrm{d} \mu
\end{align}
In the limit $\beta\widetilde{\beta}\to \infty$, we have
 \begin{align}
&\Omega= -\gamma A = -\widetilde{\beta}\int_0^{\nu}\Phi(\mu)  \mathrm{d} \mu \notag\\
 &   \frac{N}{\widetilde{\beta}}=\int_0^{\nu} \rho(\mu)\mathrm{d} \mu = \Phi(\nu)
\end{align}
where $\gamma$ is the pressure in $2D$. The above then leads to
\begin{align}
    &\frac{N}{\widetilde{\beta}} =  \frac{1}{2\pi}\left[\ln\widetilde{\beta} ~\nu-\operatorname{Im}\ln\Gamma\left(1+i\widetilde{\beta}\nu\right)\right]\notag\\
    &\gamma A = \frac{\widetilde{\beta}}{2\pi}\left[\frac{\widetilde{\beta}^2}{2}\ln\widetilde{\beta}~\nu^2-\ln\sqrt{2\pi} +\operatorname{Re}\psi^{(-2)}\left(1+i\nu\widetilde{\beta}\right)\right]
\end{align}
From the above it is easy it see that $\gamma > 0$ for a black-hole of generic mass. However, the black-hole event horizon also has surface tension which can be ascertained via standard black-hole thermodynamics \cite{Smarr:1972kt, Mazur:2015kia}
\begin{align}
    dU = \sigma dA \quad \sigma = \frac{\kappa}{8\pi}
\end{align}
where $\kappa$ is the surface gravity. Also, note that we also have
\begin{align}
    A = \frac{16\pi}{\widetilde{\beta}^2} \quad\kappa = \frac{\widetilde{\beta}}{4} 
\end{align}
Since, the black-hole event horizon is demonstrated to have an associated fermionic density of state $\rho(\mu)$, the event horizon will have $2D$ surface pressure $\gamma$ which will contribute to the first law of thermodynamics via the work term $dW = \gamma dA$. Hence, from the first law of thermodynamics, we have
\begin{align}
    &dU = dQ-dW\\
    &\implies dU = TdS -\gamma dA
\end{align}
Since, a black-hole has no interior or exterior pressure, they don't contribute to the above work term. Also, for a black hole, we have\cite{bekenstein1973black}
\begin{align}
    T = \frac{\kappa}{2\pi} \quad S = \frac{A}{4}
\end{align}
the first law then becomes
\begin{align}
    dU = \left(\frac{\kappa}{8\pi}-\gamma\right)dA =(\sigma-\gamma)dA
\end{align}
where $\frac{\kappa}{8\pi}$ is identified as the effective surface tension $\sigma$ of the black-hole event horizon \cite{Smarr:1972kt}. Notice that the pressure term only contributes to the above in the $M\to 0$ (large $\widetilde{\beta}$) limit as it should to be consistent with
standard black-hole thermodynamics. Indeed, $\gamma$ represents a correction to the first law in $M\to 0$ regimes. The remnant configuration is, therefore, achieved when the surface tension of the event horizon equals the pressure i.e.
\begin{align}
     \sigma = \gamma \label{rem_eq}
\end{align}
The $\widetilde{\beta}$ that satisfies the above for a fixed $\nu$ precisely corresponds to the remnant mass. The stability condition for a thermodynamic system is given by \cite{callen2006thermodynamics}
\begin{align}
    \frac{\partial^2U}{\partial S^2} > 0
\end{align}
Since, $S = \frac{A}{4}$, we have
\begin{align}
     dU = (\sigma-\gamma)dA \implies \frac{d U}{d S} = 4(\sigma-\gamma)
\end{align}
from which we obtain
\begin{align}
    \frac{d^2U}{dS^2} = \frac{\sigma'(M)-\gamma'(M)}{2\pi M}
\end{align}
The stability condition is satisfied when $\sigma'(M)-\gamma'(M) > 0$. At thermal equilibrium, we have $dU = 0$, which leads to
\begin{align}
    \sigma = \gamma \implies dA = 0
\end{align}
We identify $M_r$, the remnant mass obtained in Section \ref{rem_con} as the mass that satisfies the above condition. Then the stability condition implies that
\begin{align}
    \frac{d^2U}{dS^2}\bigg|_{M = M_r} = \frac{\sigma'(M_r)-\gamma'(M_r)}{2\pi M_r} > 0 \quad \sigma(M_r) = \gamma(M_r)\label{stabcon}
\end{align} 
It can be checked that the remnant masses obtained in Figure \ref{rem_sol} satisfies the above stability condition. The remnant mass is, therefore, determined upto an undetermined parameter $\nu$ which is the Fermi energy of the matrix model\footnote{See Figure \ref{rem_sol} for remnant configurations at various values of $\nu$.}. This is very reminiscent of how remnants are realized in other formalisms like GUPs and Non-commutative black-holes. In GUP, the remnant mass is predicted to be \cite{Ong:2024dnr}
\begin{align}
    M_{r} = \frac{\sqrt{\alpha}}{2} \quad \text{(GUP)}
\end{align}
where $\alpha$ is an undetermined parameter in the GUP. As one can see, $\alpha$ can be chosen in such a way as to fit any one of the remnant masses computed above. The same holds for Noncommutative Bardeen Black-Holes (NBBH) where the remnant mass is predicted to be \cite{Mehdipour_2016}
\begin{align}
     M_r = 1.9 \sqrt{\theta} \quad \text{(NBBH)}
\end{align}
where $\theta$ is an undetermined parameter related to the noncommutativity of the manifold. In the above, we specifically chose NBBH of zero magnetic charge. Just like with $\alpha$ in GUP, $\theta$ can also be chosen in such a way as to fit any of the remnant masses computed above. Notice how these parameters play an analogous role to our Fermi level $\nu$.  Coincidentally, our analysis also bears a very close resemblance to an analysis done by modelling the event-horizon as a Bose gas instead\cite{sakalli2025thermodynamic}. This could be due to a hidden Boson-Fermion duality at the event-horizon as hinted earlier which we intend to explore in a future work. 

\section{Discussion}
In this exercise, we utilized two standard techniques of Quantum Field Theory (QFT) - symmetry and resummation in order to demonstrate existence of remnant scenarios in Hawking's original computations. We utilized the emergent conformal symmetry in the $M\to 0$ limit to first demonstrate that remnants naturally arise in Hawking's formula in Eq. (\ref{mlr}). To further provide a phenomenological picture for the remnant formation, we utilized the resummation technique to demonstrate a duality between black-hole horizon and $N\times N$ matrix models. This duality with the $N\times N$ matrix models, which is essentially $2D$ fermionic gas, meant that the event-horizon has surface pressure $\gamma$ which facilitates remnant formation by opposing the effective surface tension $\sigma$. Resummation techniques are often used in QFTs to test for dualities this way. This is computationally a more natural means of demonstrating existence of remnant compared to other methods in the literature like Generalized Uncertainty Principles (GUPs), Noncommutative black-hole models and LQG-inspired white-hole remnant scenarios as these methodologies require several external assumptions about deviations from standard physics at Planck's scale to facilitate the remnant configuration. For instance, in GUPs, one assumes a modification of the conventional Heisenberg Uncertainty Principles at the Planck's scale. These modifications impose a minimum uncertainty in the position which facilitates the remnant formation. Noncommutative black-hole models are similar to GUPs, in that, as they assume additional uncertainty relations between the position coordinates that facilitate the remnant scenario. LQG-inspired white-hole remnants assume deviations from the Schwarzschild geometry near the Planck's scale to facilitate white-hole remnants. Computations like these show that simple semi-classical methods utilized by Hawking coupled with QFT techniques and symmetry arguments in quantum gravity are much more powerful and elegant than was previously understood as they can seamlessly transition into predictions and expectations of theories of quantum gravity \cite{Arraut:2025qsh}.
\section{Summary and Conclusions}
In Section \ref{rem_con}, we complemented Hawking's original computations on black-hole radiance using conformal symmetry arguments by showing emergence of remnants in the $M \to 0$ limit. The remnant mass $M_r$ appeared as an undertermined constant of integration. To estimate the remnant mass $M_r$, we utilized the resummation techniques in Section \ref{rem_MMo} to rewrite the Hawking's formula for black-hole evaporation as a Fermi's Golden Rule for transition rates in some quantum mechanical system. This quantum mechanical system turns out to be $N\times N$ matrix model which is essentially a model for $2D$ fermionic gas. From the transition rate, we extracted the density of state for the fermionic gas which allowed us to compute the surface pressure and other macrostates of the event-horizon. The mass at which the surface pressure balanced the surface tension of the event-horizon was identified with the remnant mass $M_r$. This facilitation of black-hole remnants via fermionization of the event-horizon hints at an underlying Boson-Fermion duality which we will investigate in our future studies. This is an important direction because a Boson-Fermion duality would imply that the horizon microstates can have a coherent state description in some dual framework\cite{senechal2004introduction} which, due to the correspondence principle, would imply that horizon microstates behave like classical states. This can have profound implications for the black-hole information paradox. This brings us to the next possible future direction. In this exercise, we did not discuss the information retention capacity of remnants which is an important property a remnant must posses in order to serve as a viable resolution to the black-hole information paradox. This is something that requires additional formalism that is beyond the scope of this letter and is better reserved for future.\\\\
Another important future direction in this regard is to consider black-holes in Finsler geometries. It has been suggested that Finsler geometries are closer to physical reality than Riemannian geometries as they naturally incorporate the unidirectionality of time \cite{randers1941asymmetrical}. The existence of black-hole thermodynamics also attests to this as the thermodynamical arrow of time is also unidirectional. Recent progress \cite{praveen2025role, narasimhamurthy2026influence} in this regard has demonstrated that this is not just a coincidence  and, hence, revisiting black-hole evaporation and remnant formation in these geometries is something we also intend to investigate in the future.

\section*{Acknowledgements}
AM would like to acknowledge the moral support of Hare Krishna Movement, Pune, India. IA and AM both would like to thank the organizers of workshop titled ``Quantum Gravity and Information in Expanding Universe" held at YITP, Kyoto in February, 2025 where this collaboration began. AM would like to thank his friend Rithwik Nivam for his continuous and steady support.

\appendix
\section{Flat space limit}\label{FLL}
\begin{align}
    &ds^2 = -\left(1-\frac{2M}{r}\right)dt^2+ \frac{dr^2}{1-\frac{2M}{r}}+r^2d\Omega^2_{s^2}\notag\\
    &\implies \lim_{M\to 0} ds^2 = -dt^2 + dr^2 + r^2d\Omega^2_{S^2}
\end{align}
In the $M\to 0$ limit, the black-hole metric reduces to the Minkowski metric where the massless scalar field theory is a CFT. Therefore, it is invariant under scaling $x \to \lambda x$ where $x =(t, r)$. Hence, $M\to 0$ can also be considered as a CFT limit. Under scaling, the massless scalar field $\Psi$ transforms as
\begin{align}
    \Psi(\lambda x) = \lambda^{-1}\Psi(x)
\end{align}
which means that the various parts of the radially symmetric ansatz for the field
\begin{align}
    \Psi(t, r, \theta, \phi) = \sum_{l, m}\int dp ~B_l(p) \frac{e^{-ip t}}{r}R_{p lm}(r)Y_{lm}(\theta, \phi)
\end{align}
must transform as 
\begin{align}
    R_{\frac{p}{\lambda}; l m}(\lambda r) = R_{p l m}(r) \quad B_l\left(\frac{p}{\lambda}\right) = \lambda B_l(p) \label{slaw2}
\end{align}
in the $M\to 0$ limit. The greybody factors $F(p)$, which are essentially $F(p) =\sum_{l}(2l+1)|B_l(p)|^2$, must have the following transformation
\begin{align}
    F\left(\frac{p}{\lambda}\right) = \lambda^2F(p) \label{scalgb}
\end{align}
To demonstrate that the above scaling relations is a consequence of the conformal symmetry in the $M \to 0$ limit, consider the radial part of the massless scalar field in the black-hole background which obeys
\begin{align}
\left[-\frac{d^2}{d r_*^2}+\left(1-\frac{2 M}{r}\right)\left(\frac{l(l+1)}{r^2}+\frac{2 M}{r^3}\right)\right] R_{p l m}(r)=p^2 R_{p l m}(r) \label{KGrad}
\end{align}
where
\begin{align}
   dr_{*} = \frac{dr}{1-\frac{2M}{r}}
\end{align}
Notice that $r_*$ transforms nontrivially under the scaling transformation $r\to \lambda r$. Therefore, Eq. \ref{KGrad} is not invariant under the following scaling transformations
\begin{align}
    r \to \lambda r \quad p \to \frac{p}{\lambda} \label{scaletran}
\end{align}
combined with the scaling relations of the fields in Eq. (\ref{slaw2}), hence, massless scalar field in black-hole background doesn't have conformal symmetry as expected. However, as $M \to 0$, $r_{*} \to r$ due to which Eq. \ref{KGrad} becomes
\begin{align}
    \left[-\frac{d^2}{d r^2}+\frac{l(l+1)}{r^2}\right] R_{p l m}(r)=p^2 R_{p l m}(r) \label{invdemo}
\end{align}
which is invariant under scaling transformations stated in Eq. \ref{scaletran} combined with field transformations in Eq. (\ref{slaw2}). Therefore, the various scaling relations obtained for these functions holds in this limit and is indeed a consequence of conformal symmetry in the $M\to 0$.
Using Eq. (\ref{scalgb}) in Eq. (\ref{lpT}), we obtain
\begin{align}
   f(\lambda\mu) = \lambda f(\mu)
\end{align}
Similarly, using Eq. (\ref{scalgb}) in Eq. (\ref{FT}) we also have under scaling
\begin{align}
  \varphi(\lambda \mu) =\lambda\varphi(\mu) \quad    \chi(\lambda\mu) =\lambda\chi(\mu)
\end{align}
\bibliographystyle{unsrt}
\bibliography{QGBib}

\end{document}